# An Adaptive Learning Mechanism for Selection of Increasingly More Complex Systems

Fouad Khan

WWF Luc Hoffmann Institute

fouadmkhan@gmail.com

*Abstract*

**Recently it has been demonstrated that causal entropic forces can lead to the emergence of complex phenomena associated with human cognitive niche such as tool use and social cooperation. Here I show that even more fundamental traits associated with human cognition such as 'self-awareness' can easily be demonstrated to be arising out of merely a selection for 'better regulators'; i.e. systems which respond comparatively better to threats to their existence which are internal to themselves. A simple model demonstrates how indeed the average self-awareness for a universe of systems continues to rise as less self-aware systems are eliminated. The model also demonstrates however that the maximum attainable self-awareness for any system is limited by the plasticity and energy availability for that typology of systems. I argue that this rise in self-awareness may be the reason why systems tend towards greater complexity.**

I. INTRODUCTION

One of the by-products of the revolution in information technology over the last three decades has been our enhanced capacity to visualize, model and understand complex phenomena. This has allowed us to identify and visualize key traits associated with complexity such as self-similarity [1] and recursion [2], interconnectedness of elements [3], high sensitivity to initial conditions [4], and theorize about the sources of these traits [5-9] and evolution of complex systems [10]. These developments though have not brought us much closer to eliminating widespread skepticism about either our ability to build predictive models of complex phenomena [11] or arrive at feasible mechanisms to describe the emergence and selection of such phenomena associated with complexity as human cognition [12], though some of the findings are already being incorporated in systems analysis, design and architecting [13]. It has also been shown that in clustering systems without noise reaching consensus is directly proportional to the size of group [14].

Recently however, it was demonstrated that traits associated with the human cognitive niche such as tool use and social cooperation can naturally emerge under the action of causal entropic forces [9]. Here, through a simple model, I demonstrate that even more rudimentary complex phenomena associated with human cognition such as 'self-awareness', can naturally emerge in systems in response to 'internal stimuli' as these internal stimuli eliminate less 'self-aware' systems.

Mechanisms proposed so far only look at external stimuli (for instance in the case of natural selection) for evolution of complexity. The mechanism proposed here acknowledges that drivers of evolution of complexity can be transformations internal to the system as well.

This paper presents a model that shows how internal stimuli through a proposed new mechanism leads to the selection of ever more complex systems.

The work presented here can be seen as a corollary of the good regulator theorem [15] and has been done to show the limitations other works that propose entropic measures as drivers for complexity [9] in a competitive environment but ignore internal stimuli; in the presence of which, competitive environment is not necessary for evolution of complexity.

II. MATERIALS AND METHODOLOGY

To construct the model we start with a system which is a 'good regulator' of itself [15]. It has been shown that any good regulator of a system is also a model of the system [15]. So if R is a good regulator of System S, then it is both a) internal to the system and b) a model of the system. Also for every 'real world' state the system S assumes, R (being a model of S) assumes a corresponding 'model' state. For the purposes of development of this model 'self awareness' (to be denoted by $\Delta$) now is defined as the change in internal model R with change in system S.

$$\Delta = \frac{dR}{dS} \quad (1)$$

Defined in this manner, self-awareness stops being a binary property but instead can be represented by a continuous bounded function (with values between 0 and 1). Instead of just either having or not having 'self-awareness', systems can have varying degrees of self-awareness; self-similarity for instance being one of the cruder forms (lower degree) of self-awareness. Every system can be imagined to have an internal model of itself within it, the question remains only of quantifying the degree of accuracy of that model.

Imagine now that starting from a state $S_o$, our system goes to a critical state $S_c$ at which the system ceases to exist due to internal stimuli. At state $S_o$, the internal model of the system is in state $R_o$. However, the internal model (which is also a good regulator) also has a state $R_c$ at which the system realizes the threat posed by the internal stimuli and adjusts its state before it reaches the critical state $S_c$. Any system for which the time $T_R$ taken for R to reach $R_c$ is smaller than the time $T_S$ taken for S to reach $S_c$ would have a longer time of existence compared to a system where $T_S<T_R$. This is the survival advantage that systems with higher $\Delta$ would have, given all else is equal. So, for a regulator to be good enough to provide survival advantage;

$$T_R<T_S$$

Where;

$$T_S = \frac{S_c - S_o}{\frac{dS}{dt}} \quad (2)$$

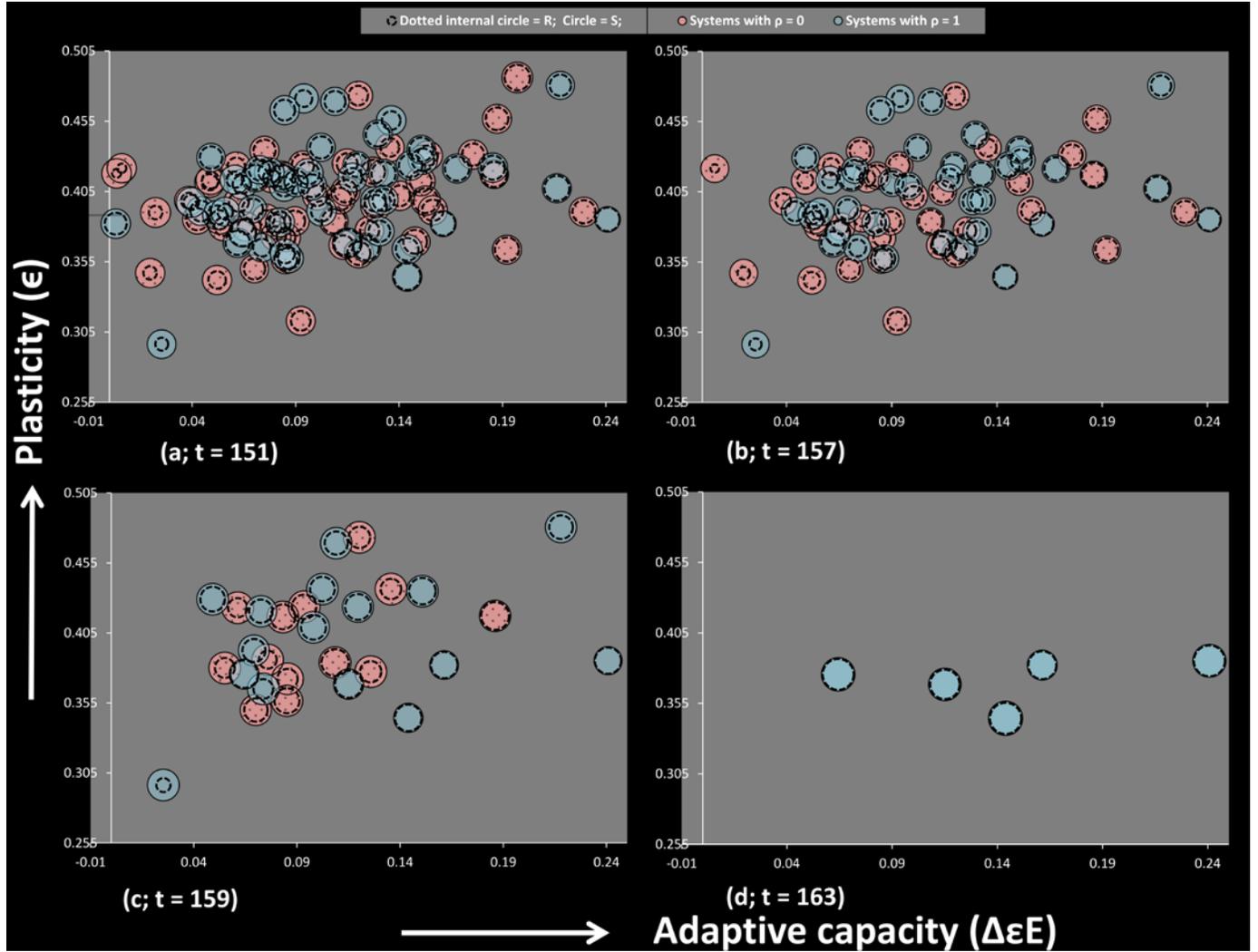

Fig. 1. Systems with lower adaptive capacity (ΔεE) die-off under adaptive selection as universe evolves over time-steps a) 151, b) 157, c) 159, d) 163. Bubbles with dotted fill are systems with agency (ρ) = 0, while bubbles with solid fill are systems with agency (ρ) = 1. Bubble size indicates value of one system state variable X. Size of the dotted outlined bubble inside bigger bubbles indicates internal model value x for the same variable X in the internal model R. As can be seen in d at time-step 163, the surviving systems are ones with very high self-awareness (dotted outline is closest to solid outline)

And

$$T_R = \frac{R_c - R_o}{\frac{dR}{dt}} \quad (3)$$

Substituting in equation 1, for an internal model to be good enough to provide survival advantage;

$$\frac{R_c - R_o}{\frac{dR}{dt}} < \frac{S_c - S_o}{\frac{dS}{dt}} \quad (4)$$

Given that $dR = \Delta dS$;

$$\frac{R_c - R_o}{S_c - S_o} < \Delta \quad (5)$$

The probability of condition specified in equation 5 being true increases with increasing $\Delta$ (where $\Delta$ is some function of the internal state variable/s of S with a range between 0 and 1) or 'self-awareness'. What this results seems to imply is that not only is a good regulator one which is a model of the system being regulated, but the better this internal model of the system is -or the higher the self-awareness of the system- the more probable it is to survive (in response to internal threats to its existence).

A simple numerical model consisting of a universe with hundred systems of varying self-awareness was built to further demonstrate how this mechanism naturally selects for systems with higher self-awareness. A binary property ρ to be called 'agency' was also introduced in the model. When R equaled $R_c$ for any system, the system readjusted only if ρ equaled 1. Overtime, we expected to see more systems with the agency switch 'on' (ρ = 1) survive as opposed to those where ρ was equal to 0. The magnitude of the readjustment depended upon the 'plasticity' of the system. Plasticity was defined as the deformation in S, per unit of available energy E, normalized to the initial value of S. Plasticity, denoted by $\epsilon$ can be expressed as;

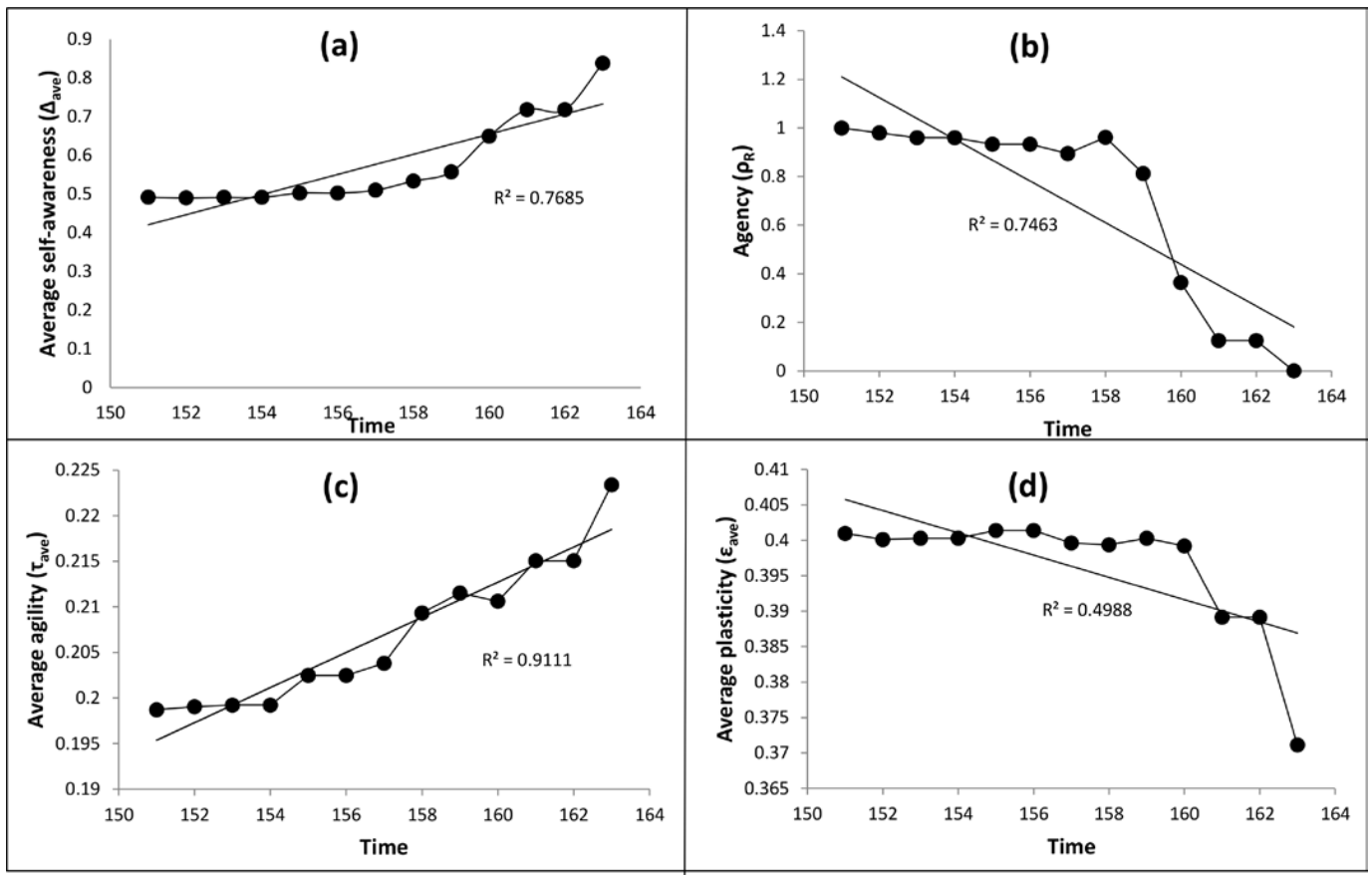

Fig. 2. Average Self-awareness of the set of living systems increases over time; b) Non-reactive systems die-off as the ratio of non-reactive to reactive systems decreases over time; c) Average agility of the set of living systems increases over time; d) Average plasticity of the set of living systems decreases over time

$$\epsilon = \frac{dS}{ES} \quad (6)$$

Further, $R_c$ depended on how quickly the system was able to identify the need for a readjustment. This property was termed 'agility'; defined as the difference between the system critical value ($S_c$) and internal model critical value ($R_c$), normalized to the system critical value $S_c$. Agility, denoted by τ can be expressed as;

$$\tau = \frac{(S_c - R_c)}{S_c} \quad (7)$$

Four parameters are monitored across the set of 'living' systems as our universe evolved and some systems were eliminated due to S having reached critical value $S_c$; i) the average self-awareness $\Delta_{ave}$; ii) ratio of number of systems with 0 agency against number of systems with agency equal to 1, $\rho_R$; iii) average agility $\tau_{ave}$ and iv) average plasticity $\epsilon_{ave}$.

### III. RESULTS AND DISCUSSION

One immediately observable fact was that all these properties across the universe evolved in bursts (spasmodically) in a manner reminiscent of scale-free networks [3].

Average self-awareness for the set of living systems was indeed seen to increase with elimination of less self-aware systems, though it was observed that the maximum attainable self-awareness for any system was limited by the product of self-awareness, plasticity and energy for that system typology. We term this product the adaptive capacity. Figure 1 shows the elimination process at four time steps during the model run.

Figure 2 shows how the monitored properties evolved over time for the universe of living systems with average self-awareness and agility increasing and ratio of positive agency over null agency systems decreasing as expected, and the average plasticity decreasing. The rise in plasticity is somewhat surprising. One should expect that the more plastic a system is, the more adaptable it should be, and hence the more resilient. What we see instead is that the systems that survive are the ones with lower plasticity.

However, from equations 1 and 6 we deduce that the change in model normalized to the original system state is equal to the product of self-awareness, plasticity and energy availability.

$$\frac{dR}{S} = \Delta \epsilon E \quad (8)$$

From equation 8 we can see that plasticity ($\epsilon$) and self-awareness ($\Delta$) are inversely related. Upon consideration this result does appear to make intuitive sense. Plasticity is a measure of how much change R can incur in S, while self-awareness is a measure of how R changes with changes in S. For any given system, the internal model can be made of either energy or matter, however in most cases, the internal model substitutes information for what is material in a system; actual quantities are replaced by say, a number representing that quantity. A state variable in the internal model say R though is more likely to either be 'information' or energy, while S, the corresponding system state variable, can be expected to have more of a material component. Imagine for instance a refrigerator, say S to a model of the refrigerator as it exists in your mind, say R. The former has a lot more material content compared to the latter. Self-awareness thus can be conceptualized as the amount of change incurred in informational content with change in real world material counterpart.

Plasticity then is a measure of how that change in information comes back and affects a change in its real world material counterpart. This loop –system affecting model affecting system- is the essence of sentience and consciousness. The term ΔϵE arrived at in equation 8 defines the upper bounds for this property for any given system. For any given system 'typology' (all systems with the same plasticity and energy availability), the product ϵE determines the upper bounds of adaptive capacity.

## IV. CONCLUSIONS

This model demonstrates not only how systems naturally tend towards greater self-awareness but also how the potential for self-awareness is restricted by the plasticity of the system and the energy availability. For any given typology (here defined by the product of plasticity and energy) thus, we will see more self-aware systems survive over longer runs, but no system can rise above the limitations imposed upon it by its typology. For planetary systems for instance, the energy available as electromagnetic forces is very weak as electromagnetic forces are weak at that scale. Energy available as gravitational force, though stronger is still comparatively weaker in terms of its ability to cause strain in the system (hence lower plasticity). This means that ΔϵE has a low value compared to organic systems where electromagnetic forces act on organic matter (much more malleable hence susceptible to higher strain and having higher plasticity). Since both ϵ and E are quantifiable terms, establishing indicative values of ϵE for different system typologies should be trivial. It could be easy to show why the organic brain with its high material malleability and energy availability offers such a generous nursery for the rise of self-awareness.

It should also be noted that for self-awareness Δ to be higher, the variables that define the state of internal model R should have higher number of stronger correlations with their corresponding counterparts in system S; the variables that define the state of system S. Higher self-awareness thus is a measure of higher number of stronger correlations between internal state variables of a system. This implies greater internal interconnectivity and thus greater complexity within the system. This means that the mechanism proposed here –an adaptive selection of better regulators- also elaborates how systems naturally tend towards higher complexity.

Since like the good regulator theorem this work is applicable to all systems from 'a cow's digestive system' [15] to national politics, examples of the mechanism proposed here can be seen in the process of regulation in many complex systems such as cities and national economies where increasing disparity and difficulty to model, increases the energy cost of regulation.

In future the research shall be expanded by empirical analysis of regulation data from complex systems such as cities and national economies.